\documentclass[pre,twocolumn,showpacs,floatfix,amsmath,amsfonts]{revtex4}

\usepackage{epsfig}
\usepackage{graphicx}
\usepackage{amsmath}
\usepackage{amssymb}
\usepackage{dcolumn}
\begin{document}

\title{Radiationless energy exchange in three-soliton collisions}

\author{Sergey V. Dmitriev$^{1}$}
\email{dmitriev.sergey.v@gmail.com}

\author{Panayotis G. Kevrekidis$^{2}$}

\author{Yuri S. Kivshar$^{3}$}

\affiliation{$^1$Institute for Metals Superplasticity Problems RAS,
Khalturina 39, 450001 Ufa, Russia
\\
$^2$Department of Mathematics and Statistics, University of
Massachusetts, Amherst, MA 01003 USA
\\
$^3$Nonlinear Physics Center, Research School of Physical Science
and Engineering, Australian National University, Canberra, 0200 ACT,
Australia }

\begin{abstract}
We revisit the problem of the three-soliton collisions in the weakly
perturbed sine-Gordon equation and develop an effective
three-particle model allowing to explain many interesting features
observed in numerical simulations of the soliton collisions. In
particular, we explain why collisions between two kinks and one
antikink are observed to be practically elastic or strongly
inelastic depending on relative initial positions of the kinks. The
fact that the three-soliton collisions become more elastic with an
increase in the collision velocity also becomes clear in the
framework of the three-particle model. The three-particle model does
not involve internal modes of the kinks, but it gives a qualitative
description to all the effects observed in the three-soliton
collisions, including the fractal scattering and the existence of
short-lived three-soliton bound states. The radiationless energy
exchange between the colliding solitons in weakly perturbed
integrable systems takes place in the vicinity of the separatrix
multi-soliton solutions of the corresponding integrable equations,
where even small perturbations can result in a considerable change
in the collision outcome. This conclusion is illustrated through the
use of the reduced three-particle model.
\end{abstract}

\maketitle

\section {Introduction}

The study of soliton collisions in nonintegrable systems
\cite{KM,Encyclo} is interesting because such systems typically
describe more realistic situations than the integrable systems where
the interactions between solitons are known to be purely elastic
\cite{AS}. In nonintegrable systems, the collision outcome can be
highly nontrivial and, depending on the degree of nonintegrability,
the collision scenario can have qualitatively different features.

For the classical $\phi^4$ equation,
\begin{equation} \label{phi4}
\frac{\partial^2 u}{\partial t^2}-\frac{\partial^2 u}{\partial x^2}
+ u -u^3 =0\,,
\end{equation}
which is rather far from an integrable system, kink collisions are
always accompanied by a  certain amount of radiation in the form of
small-amplitude wave packets, as well as by the excitation of the kink's
internal modes
\cite{IntModes}. The latter are responsible for several effects in
the $\phi^4$ kink-antikink collisions. In particular, the resonant
energy exchange between the translational motion of the kinks and
their internal modes explains the fractal kink-antikink scattering
\cite{anninos}. This is a topic that was initiated by the numerical
studies in Ref.~\cite{Campbell} (see also Ref.~\cite{belova} and
references therein), and it is still under active
investigation~\cite{goodman}.

For long time, the excitation of the soliton internal modes and the
radiation losses were thought to be two major manifestations of
inelasticity of the soliton collisions in nonintegrable models.
However, a qualitatively different manifestation was recently
identified, namely, the radiationless energy exchange (REE) between
colliding solitons
\cite{IVF,Helge,pss,Mirosh,Miyauchi,FractalBreathers,Chaos,PRE2002,PRE2003}
(abbreviated as "REE" in Ref.~\cite{Helge}, a designation that we
will use hereafter).

The energy transferred to soliton internal modes in soliton
collisions, for small $\epsilon$, is typically proportional to
$\epsilon^2$, and the same is true for the radiation losses (here
$\epsilon$ is the coefficient in front of a perturbation term, added
to an integrable equation). Terms proportional to $\epsilon^2$
appear as the lowest-order correction terms in the collective
variable approaches used to describe the soliton's internal modes
\cite{IntModes}; the kink dynamics in the discrete $\phi^4$ equation
\cite{CYip}; the kink and breather dynamics in the discrete
sine-Gordon equation (SGE) \cite{BWE,BP}; and the radiation from the
discrete SGE kink \cite{Ishimori} and from the soliton in the
discrete nonlinear Schr\"odinger equation (NLSE) \cite{KM}.
On the other hand, the degree of inelasticity due to the REE effect,
when the latter is present (see details below) grows proportionally
to $\epsilon$ \cite{Helge,pss}. This means that for weakly perturbed
integrable systems the REE effect becomes a dominant manifestation
of the inelasticity of collision, while the soliton's internal modes
and radiation become increasingly important with stronger deviations
from integrable case.

The REE effect can also be responsible for the fractal soliton
scattering which was demonstrated for the first time in
\cite{FractalBreathers} for the weakly perturbed SGE and later for
the weakly perturbed NLSE \cite{Chaos,PRE2002}. In contrast to those
studies, in Refs.~\cite{Yang,Tan} the fractal scattering of vector
solitons in the coupled NLSE was attributed to the resonance energy
exchange between the soliton's translational and internal modes,
i.e., through the mechanism similar to that operating for the
$\phi^4$ kinks \cite{anninos,Campbell,belova,goodman}.



Fractal soliton scattering in the weakly perturbed NLSE was
explained qualitatively in the frame of a very simple model
\cite{Chaos} and for the generalized NLSE in the context of a more
elaborate collective variable approach \cite{ZhuYang}, based on the
method of Karpman and Solov'ev \cite{KarpmanSolovev}. Remarkably,
the soliton's internal modes were not involved into consideration in
\cite{Chaos,ZhuYang} indicating that the underlying dominant
mechanism for the fractal scattering was the REE effect (rather than
the internal mode excitation).

For weakly perturbed integrable systems, parameters of the colliding
solitons where the REE effect is observed can be found from the
analysis of the corresponding {\em integrable} equation. This was
done for the weakly perturbed SGE in \cite{Mirosh} and for the
weakly perturbed NLSE in \cite{PRE2002} using the fact that the REE
effect is observed in the vicinity of separatrix multi-soliton
solutions of the integrable equation.

In the case of moderate deviation from integrability, it becomes
increasingly important to check if the degree of nonintegrability
and the sign of perturbation allows for the existence of
noticeable soliton internal modes before one can judge to which
extent the REE effect and the soliton internal modes contribute to
the inelasticity of collision (see, e.g., Sec. II\ D in
\cite{PRE2003}). The effect of the REE effect in the case of a
moderate degree of nonintegrability has studied far less
extensively than in the case of weak perturbation, though valuable
results have been recently obtained for the discrete NLSE
\cite{ioanna} and for the  generalized NLSE with various types of
the nonlinear term \cite{ZhuYang}, where a general system of
ordinary differential equations was derived for the velocities,
amplitudes, positions and phases of the solitary waves. The latter
was shown to qualitatively and quantitatively match the
predictions of the full model.


In this paper, we study the REE effect in three-soliton collisions
of a weakly perturbed sine-Gordon equation. In the frame of the
three-particle model, we demonstrate that the REE effect is directly
related to a separatrix solution, and it offers a very transparent
explanation of the origin of fractal soliton scattering. We also
classify in a general way the potential for emergence of such
phenomena in three-kink collisions of the weakly perturbed
sine-Gordon model.


The paper is organized as follows. In Sec. \ref{Sec:II}, the
results of numerical study of the degree of inelasticity of
three-soliton collisions in the Frenkel-Kontorova model Eq.
(\ref{FrenKon}) are presented. First, the collisions between three
kinks/antikinks are analyzed in Sec. \ref{Sec:IIA} and then the
kink-breather collisions are investigated in Sec. \ref{Sec:IIB}.
The three-particle model is introduced and analyzed in Sec.
\ref{Sec:III}. The discussion of the results and our conclusions
are presented in Sec. \ref{Sec:IV}.

\section {Three-soliton collisions in weakly discrete SGE}
\label{Sec:II}

To study the effects of non-integrability on the soliton collisions
it is desirable to have a model with tunable deviation from an
integrable case [c.f. with Eq. (\ref{phi4}) which does not have such a
parameter]. The Frenkel-Kontorova (FK) model,
\begin{eqnarray} \label{FrenKon}
   \frac{{\rm d}^{2}u_{n}}{{\rm d}t^{2}}
   -\frac{1}{h^{2}}(u_{n-1}-2u_{n}+u_{n+1})+\sin u_{n}=0\,,
\end{eqnarray}
which is a discretization of the integrable SGE,
\begin{equation} \label{SGE}
   u_{tt}-u_{xx}+\sin u =0\,,
\end{equation}
is a convenient choice for such a study
\cite{Mirosh,Miyauchi,FractalBreathers}. The (singular) perturbation
parameter in Eq. (\ref{FrenKon}) is $\epsilon=h^{2}$ (with $h$
being the lattice spacing); the lowest order correction to
SGE due to the discretization can be quantified, upon a
Taylor expansion of the second difference, as $(\epsilon/12)u_{xxxx}$.

The exact three-soliton solutions to SGE are well known
\cite{Hirota,Mirosh}. The solutions are the combinations of
single-soliton solutions, namely kinks ($K$) or antikinks ($\overline
K$), having the topological charges $q=1$ and $q=-1$,
respectively, and two-soliton solutions, namely breathers ($B$), which
are actually the kink-antikink oscillatory bound states.

Energy $E$ and momentum $P$ of one SGE kink are defined by its
velocity $V$ as follows
\begin{equation} \label{EPK}
   E_{K}=8\delta, \quad P_{K}=8V\delta,
   \quad {\rm where} \quad \delta^{-1}=\sqrt{1-V^2}.
\end{equation}
Energy and momentum of a breather are defined by its frequency
$\omega $ and velocity $V$:
\begin{eqnarray} \label{EPB}
   E_{B}=16\eta\xi ,\,\,\,P_{B}=16\eta\xi V, \nonumber \\
   {\rm where} \quad
   \xi^{-1}=\sqrt{1-V^2},\,\,\,\eta=\sqrt{1-\omega^{2}}.
\end{eqnarray}

Below we describe the numerical results for the three-soliton
collisions in the weakly discrete ($h^{2}=0.04$) SGE Eq.
(\ref{FrenKon}). The exact three-soliton solutions to SGE were
employed for setting the initial conditions. The equations of
motion Eq. (\ref{FrenKon}) were integrated with the use of the
St\"ormer method of order six. We register the parameters of
quasi-particles after their collision and compare them with those
before the collision. The larger the change in the parameters,
the more inelastic the collision is.

\subsection {Three-kink collisions}
\label{Sec:IIA}

We number the kinks in a way that at $t=0$ (before the collisions)
their initial positions are related as
$(x_{0})_{1}<(x_{0})_{2}<(x_{0})_{3}$ and momenta as
$P_{K_{1}}>P_{K_{2}}>P_{K_{3}}$.
Here we consider only symmetric collisions with
$P_{K_{1}}>0$, $P_{K_{2}}=0$, and $P_{K_{3}}=-P_{K_{1}}$.
Consideration of non-symmetric collisions does not bring any new
important physical effects. For the symmetric collisions it is
convenient to set $(x_{0})_{1}=-(x_{0})_{3}$ so that the
three-soliton collisions are expected when $(x_{0})_{2}$ is close
to the origin, otherwise the two successive two-soliton collisions
will take place. Thus, among the kink's initial positions
$(x_{0})_{i}$ the only essential parameter is $(x_{0})_{2}$.

Finally, our three-kink system is defined by the topological
charges of the kinks. There are eight possible variants in assigning the
charges to the three kinks, which, due to symmetry, can be
divided into three groups of topologically
different collisions: $K\overline{K}K=\overline{K}K\overline{K}$,
$KKK=\overline{K}\overline{K}\overline{K}$, and
$KK\overline{K}=\overline{K}KK=K\overline{K}\overline{K}=\overline{K}\overline{K}K$.
We will refer to each group by referring to their first member.

The collision outcome is presented by the momenta of kinks after
collision, $\tilde{P}_{K_{j}}$, as the functions of $(x_{0})_{2}$
for a given $P_{K_{1}}$, which defines the initial momenta of the
kinks, $P_{K_{j}}$, as described above. In some cases a
kink-antikink pair can merge into a breather. In those cases we
assumed that the kinks constituting the breather share its
momentum equally, in order to plot their momenta.

\begin{figure}
\includegraphics{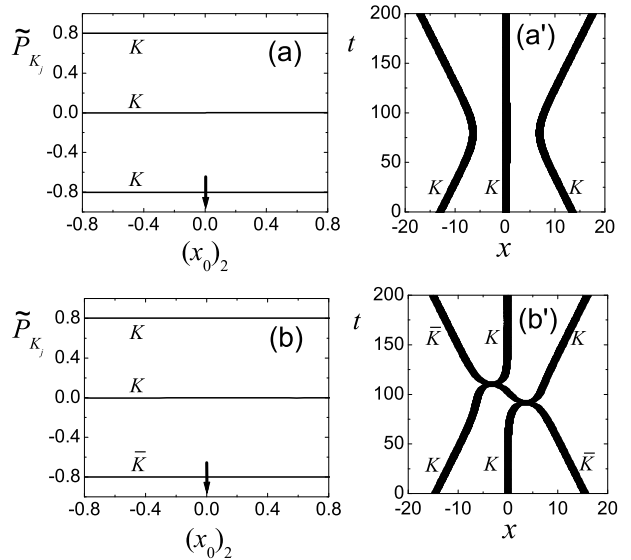}
\caption{Numerical results for (a), (a') $KKK$ and (b), (b')
$KK\overline{K}$ collisions in FK model. The left panels show the
momenta of the kinks after collision $\tilde{P}_{K_{j}}$ as the
functions of the initial position of middle kink, $(x_{0})_{2}$.
In both cases momenta of the kinks before the collision were
$P_{K_{1}}=-P_{K_{3}}=0.8$ and $P_{K_{2}}=0$ and they are nearly
same after the collision meaning that the collisions are
practically elastic for any $(x_{0})_{2}$. The right panels show the
examples of collisions on the $(x,t)$ plane for $(x_{0})_{2}=0$ by
plotting the regions with the energy density greater than certain
value, so that the cores of the solitons are clearly seen.}
\label{sfig1}
\end{figure}

\begin{figure}
\includegraphics{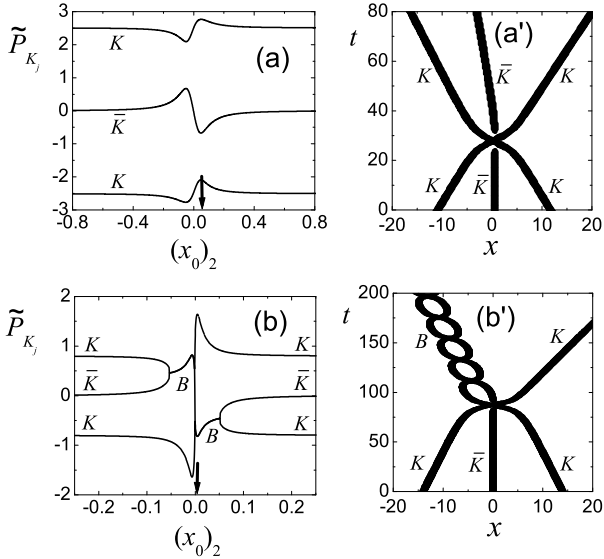}
\caption{Numerical results for $K\overline{K}K$ collisions with
(a), (a') $P_{K_{1}}=2.5$ (larger collision velocity) and (b,b')
$P_{K_{1}}=0.8$ (smaller collision velocity) in the FK model. The left
panels show the momenta of the kinks after collision
$\tilde{P}_{K_{j}}$ as the functions of the initial position of
middle kink, $(x_{0})_{2}$. Collisions are strongly inelastic for
$(x_{0})_{2}$ close to the origin. The right panels show the examples
of collisions on the $(x,t)$ plane for (a') $(x_{0})_{2}=0.05$ and
(b') $(x_{0})_{2}=0.01$ by plotting the regions with the energy
density greater than a certain value, so that the cores of the
solitons are clearly seen. Collision with a larger velocity in
(a), (a') results only in quantitative change of kink parameters
while collision with a smaller velocity in (b), (b') may result in
fusion of a kink-antikink pair in a breather.} \label{sfig2}
\end{figure}

The results for the $KKK$ collisions are shown in Fig. \ref{sfig1}
(a) for $P_{K_{1}}=0.8$. Similar results for the $KK\overline{K}$
collisions are shown in Fig. \ref{sfig1} (b) also for
$P_{K_{1}}=0.8$. The results for the $K\overline{K}K$ collisions
are shown in Fig. \ref{sfig2} (a) for $P_{K_{1}}=2.5$ (larger
collision velocity) and in Fig. \ref{sfig2} (b) for
$P_{K_{1}}=0.8$ (smaller collision velocity).

In the panels (a') and (b') of Fig. \ref{sfig1} and Fig.
\ref{sfig2} the examples of collisions are presented on the
$(x,t)$ plane by showing the regions of energy density
greater than a certain value, so that the cores of the solitons are
clearly seen. These examples are given for the particular values
of the initial coordinate of the middle kink, $(x_{0})_{2}$,
indicated by the arrows in the corresponding panels at left.

First we note that $KKK$ and $KK\overline{K}$ collisions are
always practically elastic regardless of the specifics of
$(x_{0})_{2}$ (see Fig.
\ref{sfig1}) and only $K\overline{K}K$ collisions are inelastic
for $(x_{0})_{2}$ close to the origin (see Fig. \ref{sfig2}). We
conclude that if a kink has positive or negative charge with equal
probability, then the REE in three-kink collisions can be expected
in two cases from eight.

Of particular importance is the fact that in the strongly
inelastic $K\overline{K}K$ collisions shown in Fig. \ref{sfig2}
the energy given to the kink's internal modes and to the radiation
is negligible in comparison to the energy exchange between the
quasi-particles \cite{Mirosh}. This is the main feature of REE
effect in soliton collisions.

The $K\overline{K}K$ collisions can be strongly inelastic because
in this case the cores of all three kinks can merge. Two-kink
collisions are practically elastic for the considered case of weak
perturbation, $h^2=0.04$, as it can be seen in Fig. \ref{sfig1}
(b), (b'). To explain why the two-kink collisions are elastic we
note that Eq. (\ref{FrenKon}) conserves energy and, for small
perturbation parameter $h^2$, the momentum is also conserved with
a high accuracy while the higher-order conservation laws of SGE
are destroyed by the weak discreteness.
The conservation of energy
and momentum sets two constraints on the two parameters of the
two-kink solution. A three-kink solution has one free parameter
and REE becomes possible if all three kinks participate in a
collision.

For the $K\overline{K}K$ collisions we note that the collision
with a larger velocity [see Fig. 2 (a), (a')] results only in
quantitative change of kink parameters, while collision with a
smaller velocity [see Fig. 2 (b), (b')] may result in fusion of a
kink-antikink pair in a breather. The result of $K\overline{K}K$
collisions is extremely sensitive to variations in $(x_{0})_{2}$ in
the vicinity of $(x_{0})_{2}=0$, especially for small collision
velocities.

A simple explanation of the fact that the collisions between two
kinks and an antikink are always practically elastic for
$KK\overline{K}$ and can be strongly inelastic in the case of
$K\overline{K}K$ will be offered in Sec. \ref{Sec:III}.

In the case of weak perturbation we never observed fractal
patterns in the three-kink collisions (recall that in the $\phi^4$
model such patterns can be observed even in two-kink collisions
but, as it was already mentioned, this model is far from an
integrable one), while it can be observed in the kink-breather
collisions, as discussed below, and in the breather-breather
collisions \cite{FractalBreathers}.


\subsection {Kink-breather collisions}
\label{Sec:IIB}

Without loss of generality, we assume $P_{K}+P_{B}=0$. Then we
have two parameters, the momentum $P_{B}$ and frequency $\omega$
of the breather. The outcome of the $KB$ collisions is studied as
a function of the initial separation between the kink and the
breather controlled by the initial kink position $(x_{0})_{K}$.

\begin{figure}
\includegraphics{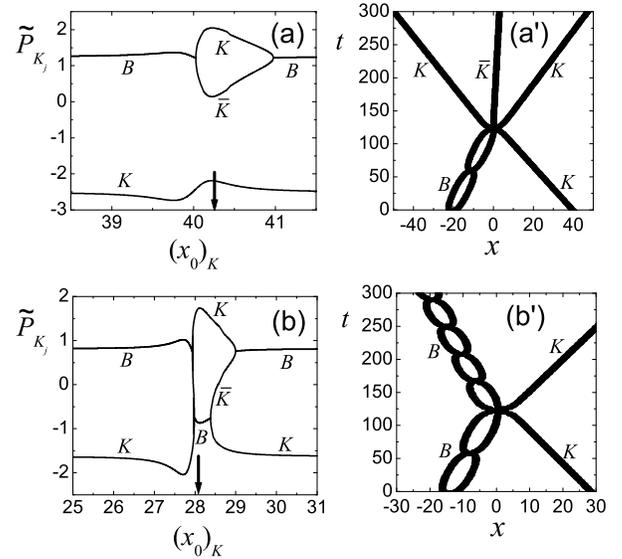}
\caption{Numerical results for the kink-breather collisions in FK
model for (a), (a') $P_{B}=2.5$ (larger collision velocity) and
(b,b') $P_{B}=1.6$ (smaller collision velocity) and $\omega=0.05$
in both cases. The left panels show the momenta of the kinks after
the collision $\tilde{P}_{K_{j}}$ as a function of the initial
position of the kink, $(x_{0})_{K}$. The right panels show
examples of collisions on the $(x,t)$ plane for (a')
$(x_{0})_{K}=40.25$ and (b') $(x_{0})_{K}=28.1$.} \label{sfig3}
\end{figure}

In Fig. \ref{sfig3} we plot the momenta of kinks after collision,
$\tilde{P}_{K_{j}}$ (including the kinks constituting the
breather, assuming as earlier that they share the breather's
momentum equally), as a function of $(x_{0})_{K}$ for (a)
$P_{B}=2.5$ (larger collision velocity) and (b) $P_{B}=1.6$
(smaller collision velocity). One can see that strong REE is
possible in the $KB$ collisions. Note that in Fig. \ref{sfig3}
only a small part of one period of the output functions is shown
for the region with strong REE. In (a) there is a range of
$(x_{0})_{K}$ where the breather obtains enough energy to split
into a kink-antikink pair [example is shown in (a')]. In (b), in
addition to this possibility, there appears a region where the
breather is reflected from the kink [example is shown in (b')].

\begin{figure}
\includegraphics{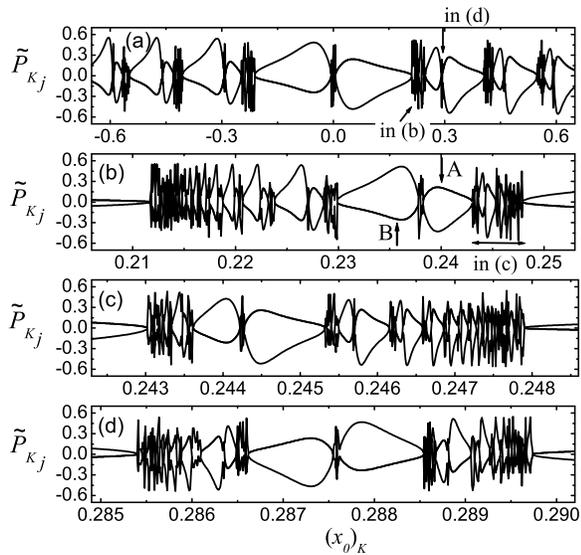}
\caption{Fractal kink-breather scattering observed for $\omega
=0.3$, $P_{B}=0$ (kink and breather have zero initial velocities).
The kinks' momenta after the collision $\tilde{P}_{K_{j}}$ are
shown as the functions of the initial position of the kink,
$(x_{0})_{K}$, at different scales. At each scale smooth regions
are separated by the apparently chaotic regions of two symmetry
types, one shown in (a) and (d) and another one in (b) and (c).
(b) and (d) present blowups of the regions indicated in panel (a);
(c) is a blowup of the region indicated in (b).} \label{sfig4}
\end{figure}


\begin{figure}
\includegraphics{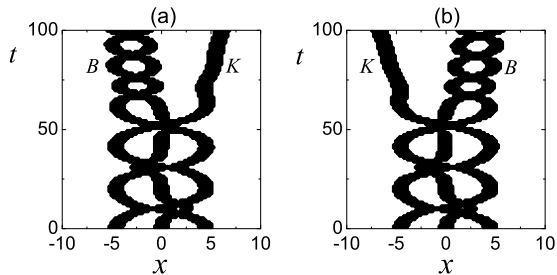}
\caption{Examples of the kink-breather dynamics for (a)
$(x_{0})_{K}=0.24$ and (b) $(x_{0})_{K}=0.236$ [indicated in Fig.
\ref{sfig4} (b) by the arrows A and B, respectively].}
\label{sfig5}
\end{figure}

\begin{figure}
\includegraphics{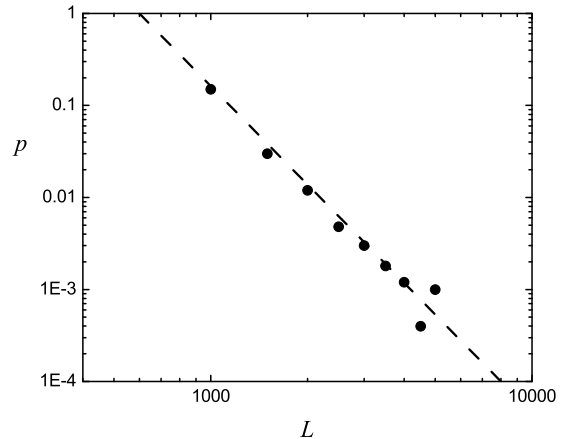}
\caption{Probability $p$ to observe the kink-breather system with
the lifetime $L$ (in Fig. \ref{sfig5} we have $L \approx 55$).
Numerical data is shown by dots and only the cases with $L>1000$
were taken into account. Dashed line is the guide for an eye and
it has slope -3.5. } \label{sfig6}
\end{figure}

For large collision velocities [somewhat larger than in Fig.
\ref{sfig3} (a), (a')] the kink passes through the breather with
no qualitative change in the collision outcome; there is only some
energy and momentum exchange between them. However, for
sufficiently small collision velocity the collision outcome as the
function of the initial separation between the kink and breather
is a fractal. An example is presented in Fig. \ref{sfig4} for
$\omega =0.3$, $P_{B}=0$ (kink and breather have zero initial
velocities), where the soliton's momenta after collision
$\tilde{P}_{K_{j}}$ are shown as the functions of $(x_{0})_{K}$.

The structure presented in Fig. \ref{sfig4} can be described as a
chain of self-similar patterns.
At each scale smooth regions are separated by the apparently
chaotic regions of two symmetry types, one shown in (a) and (d)
and another one in (b) and (c). (b) and (d) present blowups of the
regions indicated in panel (a); (c) is a blowup of the region
indicated in (b).

Two examples of the kink-breather dynamics are given in Fig.
\ref{sfig5} for (a) $(x_{0})_{K}=0.24$ and (b) $(x_{0})_{K}=0.236$
[indicated in Fig. \ref{sfig4} (b) by the arrows A and B,
respectively]. The three-particle solution has a certain lifetime
$L$ (in this example $L \approx 55$) and then it splits into a
kink and a breather. Similar dynamics has been reported, e.g., for
the breather-breather system in the weakly discrete FK model
\cite{FractalBreathers}, in the weakly perturbed NLSE
\cite{Chaos}, and recently for the generalized NLSE
\cite{ZhuYang}. Thus, this type of dynamics is rather general. For
the two-soliton collisions in the weakly perturbed NLSE we have
estimated numerically the probability $p$ to observe the
three-particle system with the lifetime $L$ and found that $p \sim
L^{-3}$ \cite{Chaos}. Here we carry out a similar estimation for
the kink-breather solution in the FK model and the result is shown
in Fig. \ref{sfig6}. The numerical data can be fitted as $p \sim
L^{-3.5}$. There is evidence that for sufficiently small frequency
of the breather the kink-breather system in the FK model with a
small $h^2$ never splits \cite{Miyauchi}.

All the important features of the $KB$ fractal scattering
including the existence of the two qualitatively different
stochastic regions in the fractal structure will be clarified in
Sec. \ref{Sec:III} with the help of the three-particle model.

\section {Three-particle model}
\label{Sec:III}

\subsection {Description of the model} \label{Sec:IIIA}

Attempting to explain the effects observed in the three-soliton
collisions in weakly perturbed SGE reported in Sec. \ref{Sec:II},
we consider the solitary waves as effective particles,
and study the dynamics of three such particles in
one-dimensional space. The particles have mass $m=8$, which is the
rest mass of SGE kink, and they carry topological charges
$q_{j}=\pm 1$. Particles with $q_{j}=1$ ($q_{j}=-1$) will be
called kinks (respectively, antikinks) by analogy with the
SGE solitons. We assume
that particles $i$ and $j$, having coordinates $x_{i}$ and
$x_{j}$, interact via the potential
\begin{equation}
U_{ij}(r_{ij}) = 16 + q_{i}q_{j}\frac{16}{\cosh (r_{ij})},\,\,\,\,\,r_{ij}=x_{j}-x_{i},\,
\label{Potential}
\end{equation}
which qualitatively approximates the interaction of two SGE kinks.
The potential of Eq. (\ref{Potential}) is attractive for
$q_{i}\neq q_{j}$ and repulsive for $q_{i}=q_{j}$.
The binding energy of the kink-antikink pair is equal to 16, which is the
energy of two standing SGE kinks. Note that for the kink and
antikink at any finite distance $r_{ij}$ the potential energy
$U_{ij}(r_{ij})$ is less than 16. If the
kinetic energy of relative motion of the particles is less than
$16-U_{ij}(r_{ij})$, then the particles cannot escape the mutual
attraction and they form an oscillatory bound state, i.e., a breather.

The Hamiltonian of the three-particle system is
\begin{equation} \label{Ham3}
   H = \frac{m}{2} \sum\limits_{j=1}^{3} v_{j}^{2}+
   U_{12}(r_{12})+U_{13}(r_{13})+U_{23}(r_{23})\,,
\end{equation}
where $v_j=dx_{j}/dt$, and there is one more integral of motion,
namely the conservation of momentum. Without loss of generality,
we assume that the total momentum in the system is equal to zero,
i.e., $m(v_{1}+v_{2}+v_{3})= 0$. Introducing new variables
\begin{equation} \label{NewVar}
   x_{2}-x_{1} \rightarrow \sqrt{3}\alpha + \beta ,\,\,\, x_{3}-x_{1}
   \rightarrow 2\beta ,\,\,\, t \rightarrow \sqrt{2m}t\,,
\end{equation}
the Hamiltonian of Eq. (\ref{Ham3}) can be presented in the form
\begin{eqnarray} \label{Ham3New}
   H = \frac{1}{2} \left(\dot{\alpha}^{2}+\dot{\beta}^{2} \right)
   +U_{12}(\sqrt{3}\alpha+\beta) \nonumber \\
   +U_{13}(2\beta)+U_{23}(\sqrt{3}\alpha-\beta)\,,
\end{eqnarray}
which is the Hamiltonian of a unit-mass particle moving
in the two-dimensional scattering potential.

Now we solve numerically three equations of motion which can be
derived from the Hamiltonian Eq. (\ref{Ham3}) and, inverting Eq.
(\ref{NewVar}), present the three-particle dynamics by the
trajectory of the particle in the ($\alpha ,\beta$)-plane.

\subsection {Separatrix three-soliton solutions to SGE} \label{Sec:IIIB}

Several separatrix three-soliton solutions to the exactly
integrable SGE Eq. (\ref{SGE}) have been reported in
\cite{Mirosh}. Here we reproduce two solutions important for our
study.

The separatrix $K\overline{K}K$ solution is
\begin{eqnarray} \label{Separatrix}
 u_{K\overline{K}K}\left( {x,t} \right) = 4\arctan (\exp x)
 + 4\arctan \frac{R}{S}, \nonumber \\
 R = \delta \left( {\sinh F - \cosh G\sinh x} \right), \nonumber \\
 S = \delta \left( {\cosh G + \sinh F\sinh x} \right) - \cosh F\cosh x, \nonumber \\
 F = - \delta x, \quad G = \delta Vt,\quad \delta^{-1} = \sqrt {1 - V^2
 }.
\end{eqnarray}
In this highly symmetric solution the anti-kink is at rest and it
is located at the point of collision of two kinks moving with the
velocities $V$ and $-V$.

\begin{figure}
\includegraphics{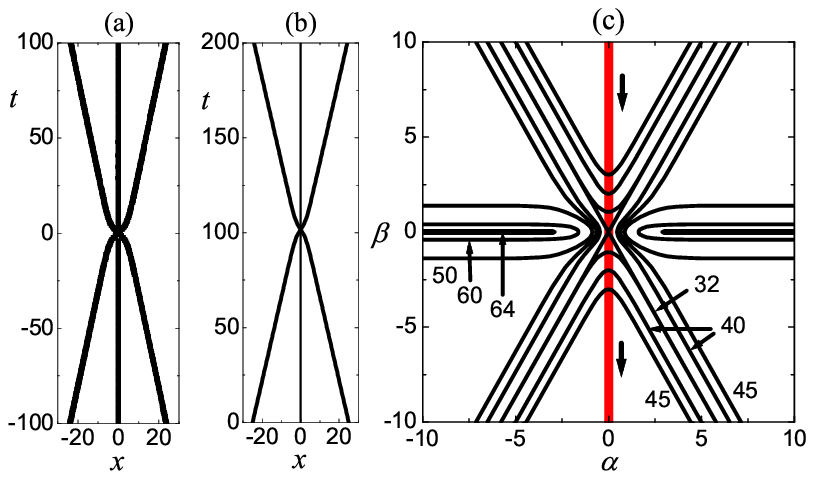}
\caption{(Color online)  (a) The SGE solution of Eq.
(\ref{Separatrix}) for $V=0.2$; (b) the three-particle dynamics
with $q_1=-q_2=q_3=1$, $(v_{0})_{1}=-(v_{0})_{3}$,
$(v_{0})_{2}=0$, $(x_{0})_{1}=-(x_{0})_{3}=-25$, and
$(x_{0})_{2}=0$ in the $(x,t)$ space; (c) the red line shows the
corresponding trajectory of the particle in the scattering
potential in the ($\alpha ,\beta$)-plane (isopotential lines are
shown in black). The particle in (c) moves along the potential
ridge and this motion is unstable. The picture in (c) gives a
visual image of the separatrix $K\overline{K}K$ solution Eq.
(\ref{Separatrix}).} \label{sfig7}
\end{figure}

\begin{figure}
\includegraphics{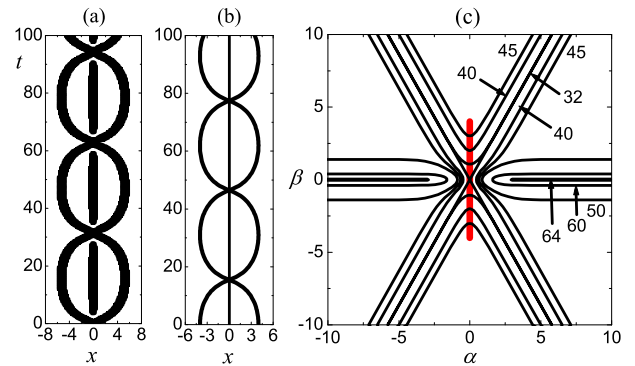}
\caption{(Color online) (a) The SGE solution Eq.
(\ref{Separatrix1}) for $\omega=0.2$; (b) the three-particle
dynamics with $q_1=-q_2=q_3=1$,
$(v_{0})_{1}=(v_{0})_{2}=(v_{0})_{3}=0$,
$(x_{0})_{1}=-(x_{0})_{3}=-4$, and $(x_{0})_{2}=0$ in the $(x,t)$
space; (c) the red line shows the corresponding trajectory of the
particle in the scattering potential in the ($\alpha
,\beta$)-plane (isopotential lines are shown in black). The
particle in (c) oscillates along the potential ridge and this
motion is unstable. The picture in (c) gives a visual image of the
separatrix $KB$ solution Eq. (\ref{Separatrix1}).} \label{sfig8}
\end{figure}

The kink-breather separatrix solution is
\begin{eqnarray} \label{Separatrix1}
 u_{KB}\left( {x,t} \right) = 4\arctan (\exp x)
 + 4\arctan \frac{X}{Y}, \nonumber \\
 X = \eta \left( {\sinh D - \cos C\sinh x} \right), \nonumber \\
 Y = \eta \left( {\cos C + \sinh D\sinh x} \right) - \cosh D\cosh x, \nonumber \\
 C = - \omega t, \quad D = \eta x,\quad \eta = \sqrt {1 - \omega ^2
 },
\end{eqnarray}
and it has only one parameter $\omega$ because it is a particular
form of the $KB$ solution where the kink and the breather have
zero velocities and zero distance between them.

In Fig. \ref{sfig7} we plot (a) the SGE solution Eq.
(\ref{Separatrix}) for $V=0.2$, (b) the three-particle dynamics in
the $(x,t)$ space for $q_1=-q_2=q_3=1$,
$(v_{0})_{1}=-(v_{0})_{3}$, $(v_{0})_{2}=0$,
$(x_{0})_{1}=-(x_{0})_{3}=-25$, and $(x_{0})_{2}=0$, and in (c)
the red line shows the corresponding dynamics in the ($\alpha
,\beta$)-plane. In (c) the isolines of the scattering potential
are also shown (black lines). The scattering potential in this
case is a superposition of a ridge along $\beta=0$ and two troughs
along the lines $\beta= \pm \sqrt{3} \alpha$. Note that the
intersection of the ridge and the two troughs forms in the
vicinity of the origin the ridge along the line $\alpha=0$; the
trajectory of the particle shown by the red line goes exactly on
the top of this ridge. Obviously, this type of motion is unstable
and, as we will see in the following, small variation in the
initial conditions may result in qualitatively different dynamics
of the particle. The picture presented in Fig. \ref{sfig7} (c)
gives a visual image of the separatrix $K\overline{K}K$ solution
Eq. (\ref{Separatrix}).

In Fig. \ref{sfig8} we plot (a) the SGE solution Eq.
(\ref{Separatrix1}) for $\omega=0.2$, (b) the three-particle
dynamics in the $(x,t)$ space for $q_1=-q_2=q_3=1$,
$(v_{0})_{1}=(v_{0})_{2}=(v_{0})_{3}=0$,
$(x_{0})_{1}=-(x_{0})_{3}=-4$, and $(x_{0})_{2}=0$, and in (c) the
red line shows the corresponding trajectory in the ($\alpha
,\beta$)-plane. The particle in (c) oscillates along the potential
ridge and, similarly to the previous example,
this motion is unstable. The picture
presented in Fig. \ref{sfig8} (c) gives a visual image of the
separatrix $KB$ solution of Eq. (\ref{Separatrix1}).

When the red line passes the origin of the ($\alpha
,\beta$)-plane, from Eq. (\ref{NewVar}) one has $x_1=x_2=x_3$,
i.e., all three particles meet at one point. In the SGE this
corresponds to simultaneous collision of all three kinks.

Looking at Fig. \ref{sfig8}(c) one can expect the possibility of
oscillation of the particle along the ridge of the scattering
potential with $\beta=0$. This is indeed possible for the
three-particle system but, from $\beta=0$ one finds from Eq.
\ref{NewVar} that $x_1(t) \equiv x_3(t)$, which cannot be realized
in the FK model because the kinks have finite width.

\subsection {Three-kink collisions} \label{Sec:IIIC}

\begin{figure}
\includegraphics{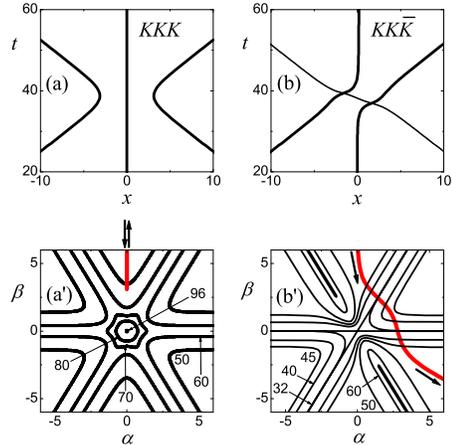}
\caption{(Color online) Comparison of (a,a') $KKK$ and (b,b')
$KK\overline{K}$ symmetric collisions. The top panels show the
three-particle dynamics in the $(x,t)$ space. The trajectories of
kinks are shown by thicker lines than those of antikinks. The bottom
panels correspondingly show the equipotential lines of the
scattering potential Eq. (\ref{Ham3New}) (black) and the
trajectory of particle (red line) in the ($\alpha ,\beta$)-plane.
The parameters are $(x_{0})_{1}=-(x_{0})_{3}=-25$, $(x_{0})_{2}=0$ and
$(v_{0})_{1}=-(v_{0})_{3}=0.6$, $(v_{0})_{2}=0$. The charges of the
particles are (a,a') $q_1=q_2=q_3=1$ and (b,b') $q_1=q_2=-q_3=1$.}
\label{sfig9}
\end{figure}

\begin{figure}
\includegraphics{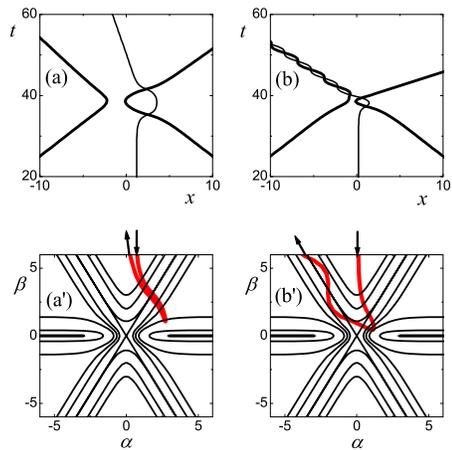}
\caption{(Color online) Sensitivity of the result of near
separatrix collision to a small deviation from $(x_{0})_{2}=0$
demonstrated by setting (a,a') $(x_{0})_{2}=1.2$ and (b,b')
$(x_{0})_{2}=0.2$. In (a,a') only a quantitative change in the
system can be seen upon collision [compare with the actual
three-kink collision in FK model shown in Fig. \ref{sfig2}(a')].
In (b,b'), kink and antikink merge in a breather [compare with
Fig. \ref{sfig2}(b')]. Other parameters:
$(x_{0})_{1}=-(x_{0})_{3}=-25$ and $(v_{0})_{1}=-(v_{0})_{3}=0.6$,
$(v_{0})_{2}=0$.} \label{sfig10}
\end{figure}

In Fig. \ref{sfig9} we compare the $KKK$ and $KK\overline{K}$
symmetric collisions in the three-particle model for
$(x_{0})_{1}=-(x_{0})_{3}=-25$, $(x_{0})_{2}=0$ and
$(v_{0})_{1}=-(v_{0})_{3}=0.6$, $(v_{0})_{2}=0$. The top panels show
the three-particle dynamics in the $(x,t)$ space. For each case,
the bottom panels correspondingly show the equipotential lines of the
scattering potential of Eq. (\ref{Ham3New}) (black) and the
trajectory of the particle (red line) in the ($\alpha
,\beta$)-plane. In (a') the scattering potential for
$q_1=q_2=q_3=1$ is a superposition of three troughs while in (b')
for $q_1=q_2=-q_3=1$ it is a superposition of a ridge and two
troughs. The potential in Fig. \ref{sfig9} (b') can be obtained
from that shown in Fig. \ref{sfig7} (c) and Fig. \ref{sfig8} (c)
through a rotation by $-\pi/3$.

In Fig. \ref{sfig9} (a), the like particles repel each other and,
in (a'), the particle hits the potential barrier and goes back. In
(b), one can see that particles collide in two successive
two-soliton collisions. In this case, the particle in (b') passes the
two potential troughs one after another and then moves away from
the origin in the direction symmetrically equivalent to the
direction it came from. Since the red line in (a') and (b') never
goes through the origin, the three particles never meet at one
point.

In Fig. \ref{sfig10} we give two examples of near-separatrix
$K\overline{K}K$ symmetric collisions in the three-particle model
for $(x_{0})_{1}=-(x_{0})_{3}=-25$ and
$(v_{0})_{1}=-(v_{0})_{3}=0.6$, $(v_{0})_{2}=0$. Recall that the
separatrix solution shown in Fig. \ref{sfig7} corresponds to
$(x_{0})_{2}=0$ but Fig. \ref{sfig10} corresponds to (a,a')
$(x_{0})_{2}=1.2$ and (b,b') $(x_{0})_{2}=0.2$. In Fig.
\ref{sfig10} (a,a'), the deviation from the separatrix is rather
large and only quantitative changes in the particle parameters can
be seen. This should be compared with the actual three-kink
collision in FK model shown in Fig. \ref{sfig2}(a'). In (b,b'),
kink and antikink merge in a breather [compare with Fig.
\ref{sfig2}(b')]. Taking into account the time reversibility in
the Hamiltonian systems, this picture can be also regarded as an
illustration of the breakup of a breather colliding with a kink.

The three-particle model explains why the REE effect is more
pronounced for the solitons colliding with a small relative
velocity. For the particle moving in the ($\alpha ,\beta$)-plane
along the separatrix [red line in Fig. \ref{sfig7}(c)], any
perturbation results in exponential in time deviation from the
potential ridge. High-speed collision results in faster passing of
the scattering potential and the trajectory of the particle cannot
be considerably changed. The situation is opposite for the slow
particle, which corresponds to the collision of solitons with a
small relative velocity.

\subsection {Kink-breather collisions} \label{Sec:IIID}

\begin{figure}
\includegraphics{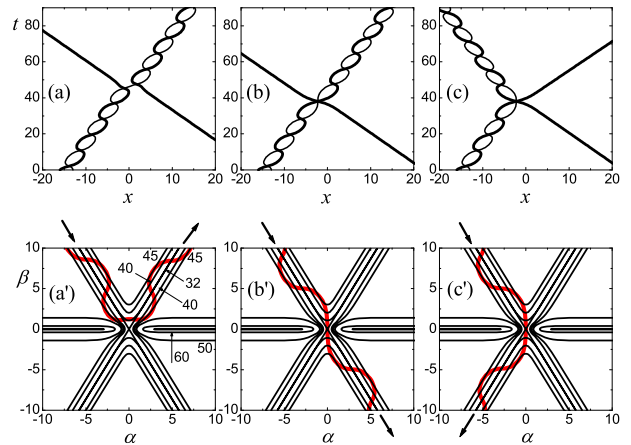}
\caption{(Color online) Three-particle model simulating the
kink-breather collisions. The top panels show the three-particle
dynamics in the $(x,t)$ space, while the bottom panels show the
corresponding trajectory of the particle in the ($\alpha
,\beta$)-plane (red line). Only the initial position of the third
particle is varied: (a) $(x_{0})_{3}=30.51$, (b)
$(x_{0})_{3}=23.398$, and (c) $(x_{0})_{3}=23.391$. Collisions in
(a),(a') and (b),(b') are elastic but in the latter case it is
close to the separatrix (see Fig. \ref{sfig8}) resulting in a
great sensitivity to variations in initial conditions, as
demonstrated in (c),(c'). The rest of the parameters are chosen
as $q_1=-q_2=q_3=1$,
$(v_{0})_{1}=(v_{0})_{2}=0.3$, $(v_{0})_{3}=-0.6$,
$(x_{0})_{1}=-16$, $(x_{0})_{2}=-13.5$.} \label{sfig11}
\end{figure}

Here we select the parameters of the three particles so as to
simulate the collisions between a kink and a breather. In particular, we
set the charges of particles as $q_1=-q_2=q_3=1$, their initial
velocities as $(v_{0})_{1}=(v_{0})_{2}=0.3$, $(v_{0})_{3}=-0.6$; the
initial positions of the particles constituting the "breather" are
$(x_{0})_{1}=-16$, $(x_{0})_{2}=-13.5$, and the third particle
initial position was varied. In Fig. \ref{sfig11} the results are
shown for (a) $(x_{0})_{3}=30.51$, (b) $(x_{0})_{3}=23.398$, and
(c) $(x_{0})_{3}=23.391$. The top panels show the three-particle
dynamics in the $(x,t)$ space, while the bottom panels show the
corresponding trajectory of the particle in the ($\alpha
,\beta$)-plane (red line).

Collisions in (a) and (b) are elastic but the difference is that
while in (a') the particle does not move along the separatrix line
$\alpha=0$, in (b') it does, and a very small change in the
initial conditions is sufficient to have a qualitatively different
result of the collision, as presented in (c),(c'), where the breather
reflects from the kink [compare (c) with actual kink-breather
collision in the FK model shown in Fig. 3(b')].

\subsection {Fractal kink-breather scattering} \label{Sec:IIIE}

\begin{figure}
\includegraphics{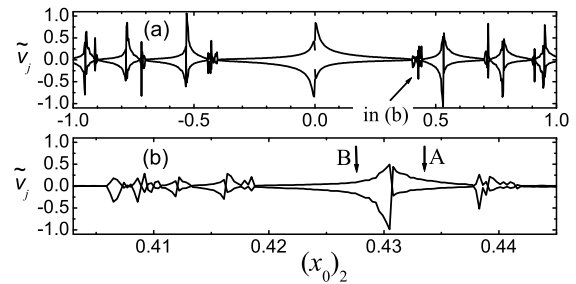}
\caption{Fractal three-particle scattering. Panels (a) and (b)
should be compared with the corresponding panels of Fig.
\ref{sfig4}. Parameters: $q_1=-q_2=q_3=1$,
$(v_0)_1=(v_0)_2=(v_0)_3=0$, $(x_{0})_{1}=-(x_{3})_{2}=-5$, and
variable $(x_{0})_{2}$.} \label{sfig12}
\end{figure}

\begin{figure}
\includegraphics{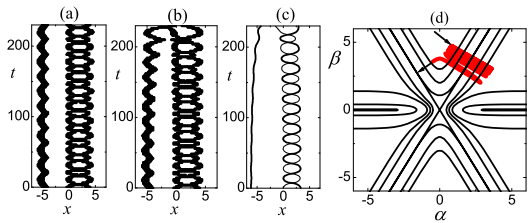}
\caption{(Color online) (a) The exact $KB$ solution to the SGE,
(b) the same solution in the weakly discrete FK model, (c) the
three-particle dynamics in the $(x,t)$ space and (d) the
corresponding dynamics in the $(\alpha,\beta)$ space. In (a), (b),
kink and breather have zero initial velocities, breather frequency
is $\omega=0.3$, and separation between the kink and the breather
is equal to $1.2$. In (c), (d), $q_1=-q_2=q_3=1$,
$(v_0)_1=(v_0)_2=(v_0)_3=0$, $(x_{0})_{1}=-6.18$, $(x_{0})_{2}=0$,
$(x_{0})_{3}=3.3$.} \label{sfig13}
\end{figure}

\begin{figure}
\includegraphics{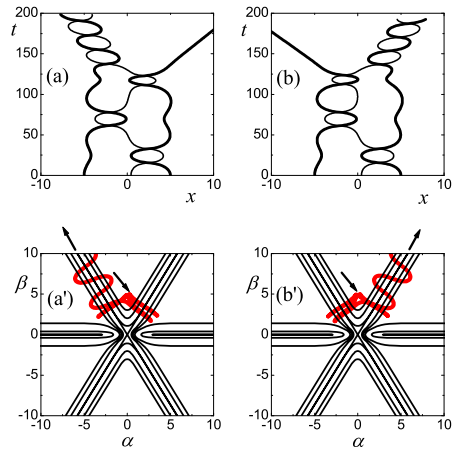}
\caption{(Color online) Illustration of one of the two possible
scenarios of fractal kink-breather scattering. (a),(b) show the
dynamics of three particles in the $(x,t)$ plane while (a'),(b')
the corresponding dynamics in the $(\alpha,\beta)$ plane. The
particle in (a'),(b') after making a few oscillations normally to
the trough $\beta=\sqrt{3} \alpha$ can cross the separatrix line
$\alpha=0$ and make a few oscillations normally to the trough
$\beta=-\sqrt{3} \alpha$ and then again cross the separatrix line
changing the trough. While the particle performs such crossings of
the separatrix it remains close to the origin and thus, all three
particles are close to each other. This defines the lifetime of
the multi-soliton bound state discussed in
\cite{FractalBreathers,Chaos,ZhuYang}. Eventually, the
particle will move away from the origin along one of the troughs.
Parameters: $(v_{0})_{i}=0$, $i=1,2,3$,
$(x_{0})_{1}=-(x_{0})_{3}=-5$, (a) $(x_{0})_{2}=0.433$ and (b)
$(x_{0})_{2}=0.4275$.} \label{sfig14}
\end{figure}

To reproduce the kink-breather fractal scattering described in
Sec. \ref{Sec:IIB} for the FK model we set the following
parameters for the particles in the three-particle model:
$q_1=-q_2=q_3=1$, $(v_0)_1=(v_0)_2=(v_0)_3=0$,
$(x_{0})_{1}=-(x_{3})_{2}=-5$, and variable $(x_{0})_{2}$.

In Fig. \ref{sfig12} we present the velocities of particles after
collision $\tilde{v}_j$ as the function of $(x_{0})_{2}$. In (b) a
blowup of the self-similar region indicated in (a) is presented.
Comparison of the panels (a) and (b) with the corresponding panels
of Fig. \ref{sfig4} reveals the qualitative similarity in the $KB$
collision outcome in the FK model and in the three-particle model.
We note that while we expect this particle model to bear the
essential qualitative characteristics of the three-particle collisions,
their details depend sensitively on the precise initial conditions;
for this reason, we expect Fig. \ref{sfig12} to match {\it qualitatively}
the results of Fig. \ref{sfig4}.

One of the important elements of the usefulness of
the three-particle model is that it gives the possibility to analyze the $KB$
fractal scattering from a different point of view, namely, by
looking at the corresponding dynamics of the particle in the
scattering potential in the $(\alpha,\beta)$ space.
In Fig. \ref{sfig13} we show (a) the exact $KB$ solution to the
SGE, (b) the same solution in the weakly discrete FK model, (c)
the three-particle dynamics in the $(x,t)$ space and (d) the
corresponding dynamics in the $(\alpha,\beta)$ space. In (a), (b),
kink and breather have zero initial velocities, breather frequency
is $\omega=0.3$, and separation between the kink and the breather
is equal to $1.2$ (we refer to the form of the $KB$ solution given
in \cite{Mirosh}). In (c), (d), $q_1=-q_2=q_3=1$,
$(v_0)_1=(v_0)_2=(v_0)_3=0$, $(x_{0})_{1}=-6.18$, $(x_{0})_{2}=0$,
$(x_{0})_{3}=3.3$.
One can see from Fig. \ref{sfig13} that the distance between the
kink and the breather does not change in time in the integrable
system [shown in (a)] but in the nonintegrable ones the distance
between them gradually decreases and they eventually collide [see
(b) and (c)]. The separated kink and breather having zero velocities
are presented in the $(\alpha,\beta)$ space by the particle
oscillating along the line normal to the trough with orientation
$\beta=\sqrt{3} \alpha$ (kink is to the left of the breather) or
$\beta=-\sqrt{3} \alpha$ (kink is to the right of the breather).
However, the troughs have a slope toward the origin of the
$(\alpha,\beta)$ plane and the oscillating particle gradually
approaches the origin, i.e., the collision point of three
particles.
After the particle has approached the origin [this situation is
shown in Fig. \ref{sfig13} (d)], two qualitatively different
scenarios giving different fractal patterns are possible.

{\em The first scenario} is shown in Fig. \ref{sfig14}. Here the
particle after making a few oscillations normally to the trough
$\beta=\sqrt{3} \alpha$ can cross the separatrix line $\alpha=0$
and make a few oscillations normally to the trough
$\beta=-\sqrt{3} \alpha$ and then again cross the separatrix line
changing the trough. While the particle performs such crossings of
the separatrix it remains close to the origin of the
$(\alpha,\beta)$ plane and thus, all three particles are close to
each other. This defines the lifetime of the multi-soliton bound
state discussed in \cite{FractalBreathers,Chaos,ZhuYang}. The
probability to have a three-particle bound state with a long
lifetime is small (see Fig. \ref{sfig6}) meaning that
eventually the particle will move away from the origin along one of the
troughs remaining in the half-plane $\beta>0$ (compare Fig.
\ref{sfig14} with Fig. \ref{sfig5} where the $KB$ dynamics in the
FK model is presented).

{\em The second scenario} is more obvious because it is directly
related to the separatrix solution Eq. (\ref{Separatrix1})
presented in Fig. \ref{sfig8}. After making a few oscillations
normally to the trough $\beta=\sqrt{3} \alpha$ as shown in Fig.
\ref{sfig13} (d), the particle can be sent by the scattering
potential almost exactly along the separatrix line $\alpha=0$.
Then the particle will make several oscillations along the ridge
of the potential, as shown in Fig. \ref{sfig8} (c), before the
inherent instability of this trajectory ``ejects'' the particle
away from the origin in one of the four directions along the
troughs $\beta= \pm \sqrt{3} \alpha$. This contrast to the first
scenario where the particle can be scattered by the potential in
the two of the four directions, namely, in the ones with
$\beta>0$.

The first scenario is associated with the self-similar regions
connecting the two ``butterflies" [see Fig. \ref{sfig4} (b) and
(c)] while the second one is associated with the self-similar
regions connecting the ``wings" of a "butterfly" [see Fig.
\ref{sfig4} (d)]. However, the whole fractal pattern is the result
of the combination of both mechanisms. Each scenario is related to
a periodic orbit of the particle in the scattering potential
\cite{Ott}. In the second scenario the periodic orbit is the
separatrix kink-breather solution shown in Fig. \ref{sfig8}, while
in the first scenario there exists an infinite set of periodic
orbits. One particular orbit can be described as follows: the
particle in the scattering potential makes $N$ oscillations almost
normally to the trough $\beta=\sqrt{3} \alpha$ and then jumps to
the trough $\beta=-\sqrt{3} \alpha$ where it also makes $N$
oscillations and then returns to the trough $\beta=\sqrt{3}
\alpha$ completing one period of the periodic orbit [Fig.
\ref{sfig14} (a') and (b') give examples when the particle makes
such jumps between the troughs $\beta=\pm \sqrt{3} \alpha$ but in
these cases the trajectories are aperiodic]. If the particle
follows a periodic orbit exactly, the three-particle system never
experiences a breakup; however, the eventual separation of the
structures is a result of the dynamical instability of such
periodic orbits.

It is well-known that the probability $p$ of the time delay $T$
for the particle interacting with the scattering potential {\em
without} the periodic orbits decreases exponentially with $T$
while in the presence of the periodic orbits it decreases
algebraically \cite{Ott}. The scattering potential in our case
does have the periodic orbits and the probability $p$ to observe a
bound state with the lifetime $L$ (analogous to the time delay
$T$) decreases algebraically, $p \sim L^{-\alpha}$. This was found
in \cite{Chaos} for the two-soliton collisions in the weakly
perturbed NLSE, and in the present study this was also confirmed
for the kink-breather system in the FK model, as presented in Fig.
\ref{sfig6}.

\section {Discussion and conclusions}
\label{Sec:IV}

Through direct numerical simulations, we have presented some of the
striking effects generated by even a weak breaking of integrability
(via discretization) in the sine-Gordon model. We have indicated
that alternative mechanisms such as the excitation of internal modes
and the emission of phonon radiation are too weak to explain the
phenomena observed in numerical simulations, and we have therefore
attributed them to the radiationless energy exchange between the
solitons. Indeed, these effects have been systematically explained
in a qualitative fashion in the framework of the three-particle
model suggested in Sec.~\ref{Sec:IIIA},
lending direct support to the conclusion that all the nontrivial
effects are due to the {\em radiationless energy exchange} between
colliding solitons
\cite{Encyclo,IVF,Helge,pss,Mirosh,Miyauchi,FractalBreathers,Chaos,PRE2002,PRE2003}.
The following is known about the REE effects: (i) Manifestations
of the REE effect grow proportionally to the perturbation
parameter $\epsilon$ while radiation and excitation of soliton
internal modes grow as $\epsilon^2$. (ii) In the sine-Gordon model
the REE effect can happen only if at least three solitons collide
simultaneously. Energy exchange in the two-soliton collision is
suppressed by the two conservation laws that remain exactly or
approximately preserved in the weakly perturbed system. (iii) The
REE effect is related to the existence of the separatrix
multi-soliton solutions to the integrable equations.
Near-separatrix motion is extremely sensitive to the perturbations
\cite{Zaslavsky}. (iv) The REE effect can be responsible for the
fractal soliton scattering.

The REE effect is generic and some of the above conclusions can be
also extended to other nearly integrable models \cite{Encyclo}. For
instance, the REE effect is observed in the weakly perturbed NLSE
already in two-soliton collisions because here each soliton has two
parameters and the total number of parameters describing the
two-soliton solution (four) exceeds the number of the remaining
conservation laws. On the other hand, the REE is {\em not} possible
in the weakly perturbed KdV equation or weakly perturbed Toda
lattice \cite{Toda} because in these cases the soliton's cores never
merge during collisions and thus, the multi-particle effects are
absent.
Interestingly, the fractal pattern of
different nature (not related to REE) is possible in KdV systems
\cite{Shapovalov}.

The three-particle model offered in the present study can be
reduced to the study of the dynamics of a particle interacting
with the two-dimensional scattering potential. Such a reduction
gives a clear interpretation of the abovementioned features of the
REE effect observed in the three-soliton collisions. Particularly,
the following features have been identified:

\begin{enumerate}

\item The three-particle model gives a visual image of the separatrix
three-kink and kink-breather solutions to the integrable SGE, see
Fig. \ref{sfig7} and Fig. \ref{sfig8}. The separatrix solution
corresponds to motion of the particle along a ridge of the
scattering potential.

\item $KKK$ collisions and $KK\overline{K}$ collisions are
always practically elastic while $K\overline{K}K$ collisions are
strongly inelastic in the vicinity of $(x_{0})_{2}$. Only in the
latter case the point moves along the ridge of the scattering
potential, which is the motion along a separatrix, see Fig.
\ref{sfig7}. Also only in $K\overline{K}K$ collisions the point
passes through the origin of the scattering potential which means that the
three kinks collide at one point simultaneously.

\item The three-particle model explains why the REE effect is more
pronounced for the solitons colliding with a small relative
velocity. For the particle moving in the ($\alpha ,\beta$)-plane
along the separatrix [red line in Fig. \ref{sfig7}(c)], any
perturbation results in exponential in time deviation from the
potential ridge. High-speed collision results in faster passing of
the scattering potential and the trajectory of the particle cannot
be considerably changed. The situation is opposite for the slow
particle, which corresponds to the collision of solitons with a
small relative velocity.

\item The fractal soliton scattering is explained by the presence of the
periodic orbits of the particle in the scattering potential.
Periodic orbits of two types were found, each of them is
responsible for a particular scenario of the particle dynamics,
and each scenario yields a self-similar pattern
for the collision outcome as a function
of a parameter, such as the location of the central effective
particle (Sec. \ref{Sec:IIIE}).

\item Periodic orbits are also responsible for the algebraic law
$p \sim L^{-\alpha}$, where $p$ is the probability to observe the
three-soliton bound state with the lifetime $L$ (see Fig.
\ref{sfig6} and Sec. \ref{Sec:IIIE}).

\end{enumerate}

In the weakly perturbed systems the REE is the dominant effect.
However, if the perturbation is not small, the energy exchange
effect is mixed with radiation and possibly with excitation of
internal modes. We thus believe that the net effect of inelasticity
of soliton collisions can be decomposed into three major parts: the
radiationless energy exchange, excitation of the soliton's internal
modes, and emission of radiation. This highlights the need for a
systematic study as a function of increasing deviations from the
integrable regime of the relative role of these three complementary
mechanisms. Such a study would be of particular interest for future
investigations.


\acknowledgments

The authors thank A.A. Sukhorukov for useful discussions. SVD
acknowledges a financial support of the Russian Foundation for Basic
Research, grant 07-08-12152. PGK acknowledges a support from
NSF-DMS-0204585, NSF-DMS-0505663, NSF-CAREER, and the Alexander von
Humboldt Foundation.

\end{document}